\documentclass[12pt,preprint2]{aastex6}

\usepackage{amsmath}
\usepackage{natbib}
\usepackage{rotating}
\usepackage{xcolor}

\newcommand{\gmass}{\log(M_*/M_{\sun})}
\newcommand{\fe}{f_{\rm esc}}
\newcommand{\sigsfr}{$M_{\sun}$~yr$^{-1}$~kpc$^{-2}$}

\slugcomment{}        

\shorttitle{}
\shortauthors{F. Bian et al.}

\begin{document}


\title{High Lyman Continuum Escape Fraction in a Lensed Young Compact Dwarf Galaxy at $Z=2.5$}


\author{Fuyan Bian\altaffilmark{1,5}, Xiaohui Fan\altaffilmark{2}, Ian McGreer\altaffilmark{2}, Zheng Cai\altaffilmark{3,6}, Linhua Jiang \altaffilmark{4} }
\altaffiltext{1}{Research School of Astronomy \& Astrophysics, Mt Stromlo Observatory, Australian National University, Canberra, ACT, 2611, Australia}
\altaffiltext{2}{Steward Observatory, University of Arizona, 933 N Cherry Ave., Tucson, AZ 85721, USA }
\altaffiltext{3}{UCO/Lick Observatory, University of California, 1156 High Street, Santa Cruz, CA 95064, USA}
\altaffiltext{4}{Kavli Institute for Astronomy and Astrophysics, Peking University, Beijing 100871, China}
\altaffiltext{5}{Stromlo Fellow}
\altaffiltext{6}{Hubble Fellow}

\begin{abstract}
We present the HST WFC3/F275W UV imaging observations of A2218-Flanking, a lensed compact dwarf galaxy at redshift $z\approx2.5$. The stellar mass of A2218-Flanking is $\gmass=9.14\substack{+0.07 \\ -0.04}$ and SFR is $12.5\substack{+3.8 \\ -7.4}$~$M_\sun$~yr$^{-1}$ after correcting the magnification. This galaxy has a young galaxy age of 127~Myr and a compact galaxy size of $r_{1/2}=2.4$~kpc. The HST UV imaging observations cover the rest-frame Lyman continuum (LyC) emission ($\sim800${\AA}) from A2218-Flanking. We firmly detect ($14\sigma$) the LyC emission in A2218-Flanking in the F275W image. Together with the HST F606W images, we find that the absolute escape fraction of LyC is $f_{\rm abs,esc}>28-57\%$ based on the flux density ratio between 1700{\AA} and 800{\AA} ($f_{1700}/f_{800}$). The morphology of the LyC emission in the F275W images is extended and well follows the morphology of the UV continuum morphology in the F606W images,  suggesting that the $f_{800}$  is not from foreground contaminants. We find that the region with a high star formation rate surface density has a lower $f_{1700}/f_{800}$ (higher $f_{800}/f_{1700}$) ratio than the diffused regions, suggesting that LyC photons are more likely to escape from the region with the intensive star-forming process.  We compare the properties of galaxies with and without LyC detections and find that LyC photons are easier to escape in low mass galaxies. 

\end{abstract}



\section{Introduction}
Observations of high-redshift quasars and galaxies and the cosmic microwave background have revealed the reionization history of the Universe, which occurred at a redshift of $z=8-9$ and largely finished at $z=6$ \citep[e.g,][]{Fan:2006lr,Fan:2006aa,Stark:2011fk,Schenker:2014aa,Bian:2015aa,Bouwens:2015ad,Robertson:2013dq,Planck-Collaboration:2016aa}. Ionizing photons emitted from high-redshift galaxies are widely considered the major sources to reionize the Universe. There are three key parameters regulating the reionization process: the total produced Lyman continuum (LyC) ionizing photons, the LyC photon escape fraction ($\fe$), and the intergalactic medium (IGM) clumping factor \citep[e.g.,][]{Robertson:2013dq}. The IGM clumping factor is difficult to measure using observations, but it can be estimated by simulations \citep[e.g.][]{Pawlik:2009aa}.The total ionizing photons produced by high-redshift galaxies can be constrained by the luminosity function and LyC photon production efficiency of high-redshift galaxies \citep[e.g.,][]{Bouwens:2015ab,Bouwens:2016ac,Livermore:2017aa}. To contribute enough ionizing photons to reionize the Universe, it is required that the LyC escape fraction ($\fe$) is comparable to or larger than $\fe=0.2$ at the epoch of the reionization \citep[e.g.,][]{Ouchi:2009aa,Robertson:2013dq}. However, the fraction of the total LyC photons that can escape from galaxies has not been well measured.

Due to the large optical depth of IGM, it is very difficult to measure the LyC escape fraction in galaxies at $z>4$. Extensive observations have been carried out to search for LyC leaking in nearby galaxies using the far ultraviolet spectroscopy \citep[e.g.,][]{Leitet:2013aa,Leitherer:2016aa,Izotov:2016aa,Izotov:2016ab} and to search for high-redshift galaxies at $z=1-4$ using deep blue optical spectroscopy \citep[e.g.,][]{Steidel:2001aa,Shapley:2006kx,Shapley:2016aa,Nestor:2013aa} and/or narrow/intermediate/broad-band UV imaging \citep[e.g.,][]{Siana:2007lr,Siana:2015aa,Vanzella:2010aa,Nestor:2011ab,Cooke:2014aa,Rutkowski:2016aa,Vasei:2016aa,Naidu:2016aa}. However, accurately measuring the escape fraction remains difficult. Most of the studies on the escape fraction in high-redshift star-forming galaxies have null detection and can only obtain upper limits ($\fe< 2\%-20\%$) even with deep stacked UV images. Meanwhile, escape fraction in high redshift galaxies may be overestimated, due to foreground contamination, especially for results based on the ground-based, seeing-limited observations \citep[e.g.,][]{Vanzella:2012aa}. Therefore, it is crucial to carry out deep, high spatial resolution imaging follow-up observations to rule out potential foreground interlopers. To date, there are only a few convincing detections of LyC emission in galaxies at $z\sim3$, including $ion2$ \citep{Vanzella:2016ab} and Q1549-C25 \citep{Shapley:2016aa}. 

To address the above problems, we study the LyC escape fraction in a sample of gravitationally lensed galaxies at $z\sim2-3$ using the HST WFC3 UV imaging observations. The high-spatial-resolution HST images rule out foreground contamination and put strong constraints on the LyC escape fraction in these lensed galaxies. In this Letter, we report a high LyC escape fraction in one of the galaxies in our HST program, a lensed compact dwarf galaxy, A2218-Flanking at $z=2.518$. Throughout this paper, the following cosmological parameters are used: Hubble constant $H_0=70$~km~s$^{-1}$~Mpc$^{-1}$, dark matter density $\Omega_{\rm M}=0.30$, and dark energy density $\Omega_\Lambda=0.70$ for a flat universe. The distance is given in physical distance. All the magnitudes are expressed in the AB magnitude system.

\begin{figure}[]
\begin{center}
\includegraphics[width=1.0\columnwidth]{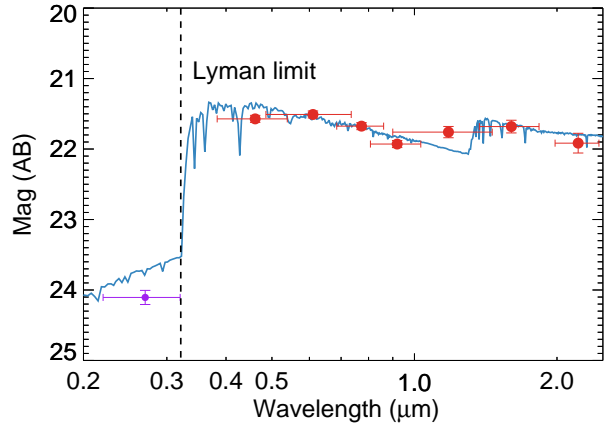}
\caption{Spectral energy distribution (SED) fitting result of A2218-Flanking. The blue curve represents the best-fit stellar synthesis model with a constant star formation history. We do not apply the interstellar medium and IGM extinction correction for the blue curve. The red points are the dereddened photometry data based on the $E(B-V)=0.11$, which is derived from the SED fitting. The purple data point represents the photometry data from HST WFC3/F275W band, which covers the blueward of the Lyman limits of A2218-Flanking. The data point has been corrected for IGM attenuation with $\exp(-\tau_{\rm IGM,800})=0.73$ (see section 3.2 for more detail). The dashed the line represents the Lyman limit at $z=2.518$.}
\label{fig:sed}
\label{fig:filter}
\end{center}
\end{figure}


\begin{figure*}[]
\begin{center}
\includegraphics[width=1.5\columnwidth]{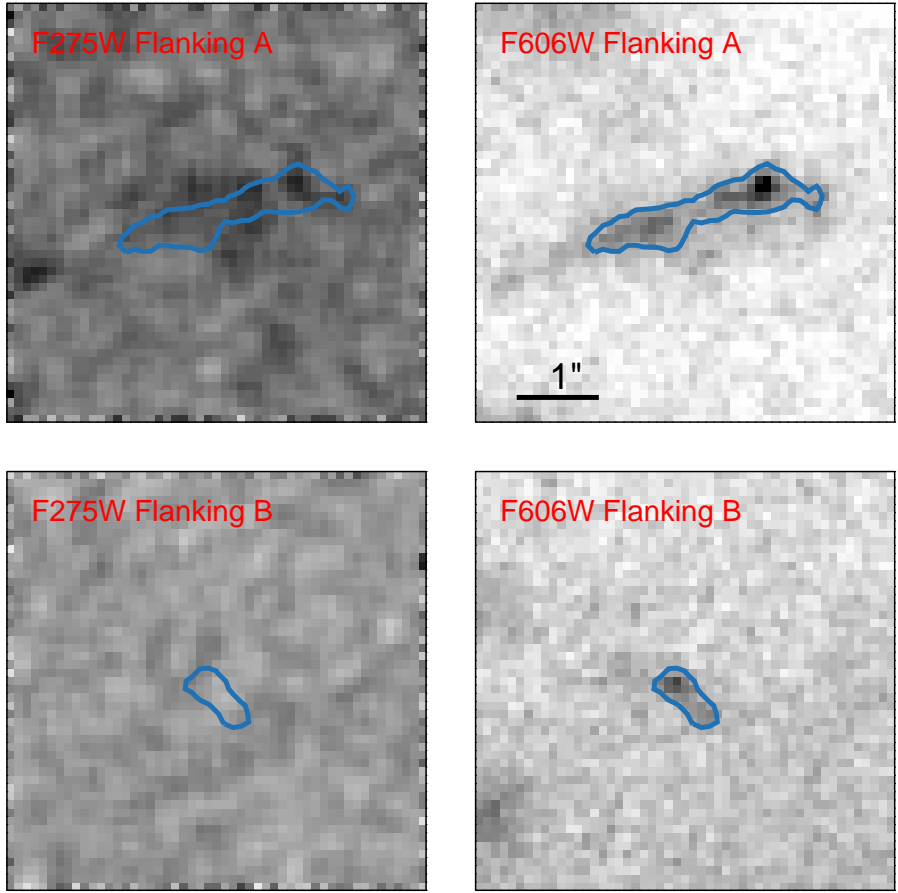}
\caption{Top panel: F275W (left) and F606W (right) images of the A component of A2218-Flanking. 
Bottom panel: F275W (left) and F606W (right) images of the B component of A2218-Flanking. 
The blue contours represent the detection region defined by the surface brightness
3$\sigma$ above the background in F606W images. North is up, and east is left.
For clarity, we smooth the F275W images by $0\farcs3$.}
\label{fig:image}
\end{center}
\end{figure*}

\section{Observations and Data Reduction}
A2218-Flanking is a compact dwarf galaxy at the redshift of $z=2.518$, which is lensed by the Abell 2218 galaxy cluster \citep{Richard:2011aa}. 
Here we summarize the properties of A2218-Flanking based on the Keck/NIRSPEC near-IR spectroscopic and HST imaging observations \citep{Richard:2011aa}. A2218-Flanking consists of two components. The total magnification factor for the two components is $13\pm2.5$. 

The size of A2218-Flanking is compact, with a half-light radius of $r_{1/2}=2.36\pm0.55$ kpc on the source plane. The gas-phase oxygen abundance in A2218-Flanking is less than 20\% of the solar abundance ($12+\log(\rm{O/H}) < 8.05$), based on the N2 metallicity indicator \citep{Pettini:2004qe}. We refer the readers to \citet{Richard:2011aa} for more detail. 

We carry out UV imaging observations of A2218-Flanking using the HST/WFC3 UVIS channel in the F275W band to cover the LyC emission (Figure~\ref{fig:filter}). The F275W band covers the observed-frame wavelength from 2200{\AA} to 3200{\AA}, which corresponds to the rest-frame wavelength of 625-910{\AA} at $z=2.518$ to ensure that there are no photons from redward of LyC (912{\AA}) to contaminate the LyC measurements. This observation is part of HST proposal program with ID 13349 (PI: X. Fan). In this program, we study the LyC escape fraction in seven lensed star-forming galaxies at $z=2.0-2.5$, including A2218-Flanking, A2218-Ebbels, CLONE,  MACS0451, J0900+2234, J0901+1814, and J1343+4155. A2218-Flanking is the only galaxy detected in the LyC images. A total of three orbits (9538 seconds) exposure on A2218-Flanking are obtained. The observations contain a six-point dithering sequence. To solve the WFC3/UVIS low charge transfer efficiency issue, we apply a background of 12$e^-$/pixel to each image to preserve faint signals and place A2218-Flanking at the bottom-left corner of WFC3/UVIS chip2, which is close to the WFC3/UVIS detector amplifier.

We use the HST WPFC2/F606W (ID: 7343, PI: Gordon Squires) image as the reference image to determine the position and detection region for A2218-Flanking. We use bright stars to align the F275 image to the F606W image, after reducing the individual F275 imaging data with the standard WFC3-UV data reduction pipeline. We combine the individual images in F275W and F606W using {\tt Astrodrizzle} 2.1.3 with a final pixel scale = 0.0396 arcsec/pixel and a final pixfrac = 0.8. Figure~\ref{fig:image} shows the A and B components of A2218-Flanking in the F275W and F606W images. 

\section{Results}
\subsection{Stellar population of A2218-Flanking}\label{stellar_population}
We use the Hyperz program \citep{Bolzonella:2000qy} to fit the \citet{Bruzual:2003aa} stellar synthesis models to the spectral energy distribution (SED) of A2218-Flanking, including broadband photometry of $HST$ F475W ($B$), F606W ($V$), F775W ($I^\prime$), F850LP, F110W ($J$), F160W ($H$) bands, and the ground-based $K$ band, which are adopted from \citet{Richard:2011aa}. We use the stellar synthesis models with a Salpeter Initial Mass Function \citep[IMF;][]{Salpeter:1955aa}, a 0.2 solar metallicity ($Z=0.004$), and a continuous star formation history with constant SFR. We apply a Small Magellanic Cloud (SMC) dust extinction curve \citep{prevot:1984aa} rather than the commonly used \citet{Calzetti:2000aa} dust extinction law, because the latter one results in an unphysically young galaxy star formation age ($t_{SF}<10$~Myr) for A2218-Flanking. \citet{Reddy:2010aa} suggested that the SMC-like extinction curve may better describe this type of galaxies. 

Figure~\ref{fig:sed} shows the SED fitting results. We find that the galaxy age is $127\substack{+33\\-14}$~Myr and the dust extinction is $E(B-V)=0.11$. After corrected for the lensing magnification ($\mu= 2.78$ mag), the stellar mass is $\gmass=9.14\substack{+0.07 \\ -0.04}$, which is about an order of magnitude smaller than that in typical star-forming galaxies at $z\sim2-3$, and the SFR is $12.5\substack{+3.8 \\ -7.4}$~$M_\sun$~yr$^{-1}$. The star formation rate (SFR) derived from the SED fitting is consistent with the dust-corrected UV-based SFR, which is $11.0\pm0.1$~$M_\sun$~yr$^{-1}$ but is higher than that in \citep{Richard:2011aa} by a factor of two. \citet{Richard:2011aa} used the dust-corrected H$\alpha$ luminosity from slit spectroscopy to estimate the SFR. The H$\alpha$-based SFR may be underestimated due to the slit loss and cause this discrepancy. Thus, we adopt the SFR derived from the SED fitting for our further analysis. 
In the fitting process, we do not take the emission line into account, which results in a slightly larger stellar mass ($\sim0.1$dex) and older galaxy age, but this does not change our main conclusions.

\subsection{LyC escape fraction}
We measure the flux densities of A2218-Flanking in the F275W and F606W images. The F275W band covers the wavelength blueward of LyC at 800{\AA}, and the F606W band samples the rest-frame UV wavelength at 1700{\AA}.  We use the F606W image as a reference image to define the detection region, in which we will  measure the flux in the F275W and F606W images. We consider the pixels with a surface brightness $3\sigma$ higher than the background in the F606W image as the detection region. Figure~\ref{fig:image} shows the detection regions for the A and B components of A2218-Flanking.

We measure the F275W and F606W fluxes in the detection regions (Figure~\ref{fig:image}). For the A and B components together, we find a total F275W band flux density of $f_{800}=0.36\pm0.03\mu$Jy and a F606W band flux density of $f_{1700}=1.83\pm0.01\mu$Jy, which leads to the ratio of $f_{1700}/f_{800}= 4.9\pm0.3$. We also measure the flux densities in the A and B components separately. For the A component, we find the F275W band flux density of $f_{800}=0.36\pm0.03\mu$Jy and the F606W band flux density of $f_{1700}=1.67\pm0.01\mu$Jy, which leads to the ratio of $f_{1700}/f_{800}= 4.6\pm0.3$. For the B component, we find a F606 band flux density of $f_{1700}=0.154\pm0.002\mu$Jy, which is about 10 times fainter than that in the A component. The F275W flux is $f_{800}=-0.006\pm0.008$. The $5\sigma$ upper limit of the F275W flux for the B component is $f_{800}<0.039\mu$Jy. If we assume that the B component holds the same $f_{1700}/f_{800}$ ratio as the A component, the expected F275W flux density of B component will be $f_{800}=0.033\mu$Jy, which is consistent with the fact that the B component is not detected in the F275W image in 5$\sigma$. These well-detected LyC fluxes suggest that A2217-Flanking has a high LyC escape fraction. 

We estimate the relative LyC escape fraction $f_{\rm esc,rel}$ by the following equation:
\begin{equation}
f_{\rm esc,rel}= \frac{L_{1700}/L_{800}}{f_{1700}{/f_{800}}}\times\exp(\tau_{\rm IGM,800}),
\end{equation}
where $L_{1700}/L_{800}$ is the intrinsic luminosity density ratio between 1700{\AA} and 800{\AA}, and $f_{1700}/f_{800}$ is the observed flux density ratio between 1700{\AA} and 800{\AA}. $\tau_{\rm IGM,800}$ is the IGM opacity at 800{\AA} at $z=2.5$ \citep[e.g.,][]{Madau:1995lr}. 

The absolute LyC escape fraction $f_{\rm esc,abs}$ is:
\begin{equation}
f_{\rm esc,abs}=f_{\rm esc,rel}\times10^{-0.4\times A_{1700}}
\end{equation}
where, $A_{1700}$ represents the dust extinction at $1700${\AA}. 

It is difficult to measure the LyC escape fraction based on one galaxy at  $z=2.5$ accurately due to the uncertainties of the intrinsic luminosity ratio, $L_{1700}/L_{800}$  and  IGM opacity, $\tau_{\rm IGM,800}$. First, the intrinsic $L_{1700}/L_{800}$  ratio is very sensitive to galaxy star formation history and galaxy age \citep{Siana:2007lr}. We find the intrinsic $L_{1700}/L_{800}=6$ based on the galaxy star formation history and galaxy age derived in section~\ref{stellar_population}. To study how the intrinsic $L_{1700}/L_{800}$ ratio changes with star formation history, we also fit the broadband SED with the stellar synthesis models with the exponential declined SFR history ($SFR\propto\exp(t/\tau)$) by changing $\tau$ from 0.1~Gyr to 1~Gyr. We find that the exponential declined SFR history usually results in a younger galaxy age (e.g., $91\substack{+11\\-38}$~Myr for $\tau=0.1$~Gyr), and the intrinsic $L_{1700}/L_{800}$ ratio can be as low as 3.5. Second, It is impossible to determine the IGM opacity for a given random light of sight. We carry out a Monte Carlo (MC) simulation of IGM absorption.  We simulate absorption systems in 10,000 lines of sight. These absorption systems follow the distributions of neutral hydrogen column density, redshift, and Doppler velocity in \citet{Inoue:2008aa}. We calculate the probability distribution of the IGM transmission in the F275W band for a random line of sight at $z=2.518$ and find that $\exp(-\tau_{\rm IGM,800})=0.22\substack{+0.51 \\ -0.22}$. By adopting $\exp(-\tau_{\rm IGM,800})<0.73$, we find that the absolute LyC escape fraction is $f_{\rm esc,abs}>57\%$ and  $f_{\rm esc,abs}>28\%$ for the intrinsic $L_{1700}/L_{800}=6$ and $L_{1700}/L_{800}=3.5$, respectively.

\subsection{Morphology of Lyman Continuum}\label{sec:morph}
The A2218-Flanking A component is well detected in the F275W image. We find that the A2218-Flanking A in the F275W image elongates to the same direction of that in the F606W image. This similarity of the morphology in the F275W and F606W images indicates that the F275W flux is not from the foreground interlopers. Unfortunately, the F275W image is not deep enough to carry out a detailed morphology analysis for the diffuse part of the galaxy. With the high spatial resolution of the HST images, we study how the $f_{1700}/f_{800}$ ratio changes across A2218-Flanking. We study a bright knot on the right (west) side of the A component (Figure~\ref{fig:image}), which is well detected in both F275W and F606W images. We measure the F275W and F606W flux density from the bright knot and find that $f_{800}=0.0343\pm0.005\mu$Jy,  $f_{1700}=0.123\pm0.002\mu$Jy, which leads to  $f_{1700}/f_{800}= 3.6\pm0.5$. This $f_{1700}/f_{800}$ ratio is significantly lower than those found in the total of the lensed galaxy and A component, which suggests that the region with higher SFR surface density has a higher escape fraction. 

\section{Discussion}
\subsection{Properties of LyC Leaking Galaxies}
It is essential to establish the relation between the LyC escape fraction and galaxy properties. This relation is crucial to predicting the LyC escape fraction of galaxies at the epoch of reionization, which cannot be directly measured \citep[e.g.,][]{Faisst:2016aa}. Based on the confirmed LyC leaking galaxies at both low and high redshifts, studies try to connect the LyC escape fraction to varieties of galaxy properties, including dust extinction (reddening), interstellar absorption lines, Ly$\alpha$ emission lines, the flux ratios between [\ion{O}{3}]$\lambda\lambda$4959,5007 and [\ion{O}{2}]$\lambda$3727, specific SFR (sSFR$=\rm{SFR}/M_{\star}$), SFR surface density \citep[e.g.,][]{Heckman:2011aa,Nakajima:2014aa,Reddy:2016aa,Izotov:2016ab,Verhamme:2017aa}. Actually, the above galaxy properties are highly correlated. For example, \citet{Bian:2016aa} found that the ionization parameter (i.e., [\ion{O}{3}]$\lambda\lambda$4959,5007/[\ion{O}{2}]$\lambda$3727) increases significantly with increasing sSFRs and decreasing galaxy sizes in star-forming galaxies, and this trend may become stronger for high-redshift galaxies. 

Therefore, it is crucial to understand the physical process behind these galaxy physical properties that regulates the LyC escape process in galaxies. \citet{Heckman:2011aa} suggested that the extreme feedback caused by the high SFR surface density plays an important role in creating channels in the interstellar medium and enabling ionizing photons to escape from galaxies. \citet{Verhamme:2017aa} did find a strong correlation between the LyC escape fraction and SFR surface density in low-redshift LyC emitters. This trend is consistent with our finding that a higher escape fraction is associated with the region with a higher SFR surface density in A2218-Flanking. 

We compare the SFR surface density in A2218-Flanking with those in the other six lensed galaxies in this HST program. We find that the SFR surface densities are comparable in these seven galaxies ($\Sigma_{\rm SFR}\sim1$~{\sigsfr}). However, only A2218-Flanking has a high LyC escape fraction, and we do not detect significant LyC flux in the remaining six galaxies, suggesting that the LyC escape is less than 5\%, assuming an average IGM optical depth for a given redshift. It is quite unlikely that the IGM transmission is an order of magnitude lower than average in all six galaxies. This piece of evidence suggests that, besides SFR surface density,  there are also other properties regulating the LyC escape fraction. One unique property of A2218-Flanking is that it has an order of magnitude lower stellar mass compared to the rest of six galaxies \citep{Richard:2011aa,Bian:2010vn}. As it is suggested in the cosmological zoom-in simulations, the LyC escape fraction is higher in lower-mass dark matter halos \citep[e.g.,][]{Kimm:2014aa}, which presumably host galaxies with lower stellar mass. 

It is worth noting that we cannot draw exclusive conclusions based on this small sample of galaxies because of the large systematic uncertainties of the $f1700/f800$ ratio and the unknown IGM optical depth for a random line of sight. Furthermore, the escape fraction may also depend on viewing angles, i.e., the ionizing radiation may be able to escape from galaxies only in some preferential directions \citep[e.g.,][]{Cen:2015aa}. Therefore, a large sample ($\sim100$) of the galaxies is required to study how the LyC escape fraction changes with galaxy properties.

\section{Conclusion}
In this Letter, we report a large Lyman Continuum escape fraction ($>57\%$) in a compact dwarf galaxy, A2218-Flanking.
Our main results can be summarized as follows:
\begin{enumerate}
\item We carried out deep HST/WFC3 F275W UV imaging observations on A2218-Flanking. The F275W band covers the Lyman continuum emission at $\sim800${\AA}. 
\item We detect significant Lyman continuum emission from A2218-Flanking. From the flux density ratio between 1700{\AA} and 800{\AA} ($f_{1700}/f_{800}$), we find
that the absolute Lyman continuum escape fraction is $f_{\rm abs,esc}>0.57$.
\item We find that the region with a higher SFR surface density has a significantly lower $f_{1700}/f_{800}$ ratio, suggesting a higher LyC escape in this region. This suggests that 
SFR surface density plays an important role in regulating the LyC escape fraction. 
\item By comparing A2218-Flanking with other galaxies in this HST program without LyC flux detections, we find that the stellar masses of galaxies may also play roles in the LyC escaping process; the LyC escape is higher in galaxies with lower stellar mass.
\end{enumerate}

\acknowledgments
We are grateful to the anonymous referee for a constructive report. We are thankful for the support for program \#13349 provided by NASA through a grant from the Space Telescope Science Institute,  which is operated by the Association of Universities for Research in Astronomy, Inc., under NASA contract NAS 5-26555.



{\it Facilities:} \facility{HST/WFC3}

\clearpage






\begin{thebibliography}{}
\expandafter\ifx\csname natexlab\endcsname\relax\def\natexlab#1{#1}\fi

\bibitem[{{Bian} {et~al.}(2016){Bian}, {Kewley}, {Dopita}, \&
  {Juneau}}]{Bian:2016aa}
{Bian}, F., {Kewley}, L.~J., {Dopita}, M.~A., \& {Juneau}, S. 2016, \apj, 822,
  62

\bibitem[{{Bian} {et~al.}(2010){Bian}, {Fan}, {Bechtold}, {McGreer}, {Just},
  {Sand}, {Green}, {Thompson}, {Peng}, {Seifert}, {Ageorges}, {Juette},
  {Knierim}, \& {Buschkamp}}]{Bian:2010vn}
{Bian}, F., {Fan}, X., {Bechtold}, J., {et~al.} 2010, \apj, 725, 1877

\bibitem[{{Bian} {et~al.}(2015){Bian}, {Stark}, {Fan}, {Jiang}, {Cl{\'e}ment},
  {Egami}, {Frye}, {Green}, {McGreer}, \& {Cai}}]{Bian:2015aa}
{Bian}, F., {Stark}, D.~P., {Fan}, X., {et~al.} 2015, \apj, 806, 108

\bibitem[{{Bolzonella} {et~al.}(2000){Bolzonella}, {Miralles}, \&
  {Pell{\'o}}}]{Bolzonella:2000qy}
{Bolzonella}, M., {Miralles}, J.-M., \& {Pell{\'o}}, R. 2000, \aap, 363, 476

\bibitem[{{Bouwens} {et~al.}(2015{\natexlab{a}}){Bouwens}, {Illingworth},
  {Oesch}, {Caruana}, {Holwerda}, {Smit}, \& {Wilkins}}]{Bouwens:2015ad}
{Bouwens}, R.~J., {Illingworth}, G.~D., {Oesch}, P.~A., {et~al.}
  2015{\natexlab{a}}, \apj, 811, 140

\bibitem[{{Bouwens} {et~al.}(2016){Bouwens}, {Smit}, {Labb{\'e}}, {Franx},
  {Caruana}, {Oesch}, {Stefanon}, \& {Rasappu}}]{Bouwens:2016ac}
{Bouwens}, R.~J., {Smit}, R., {Labb{\'e}}, I., {et~al.} 2016, \apj, 831, 176

\bibitem[{{Bouwens} {et~al.}(2015{\natexlab{b}}){Bouwens}, {Illingworth},
  {Oesch}, {Trenti}, {Labb{\'e}}, {Bradley}, {Carollo}, {van Dokkum},
  {Gonzalez}, {Holwerda}, {Franx}, {Spitler}, {Smit}, \&
  {Magee}}]{Bouwens:2015ab}
{Bouwens}, R.~J., {Illingworth}, G.~D., {Oesch}, P.~A., {et~al.}
  2015{\natexlab{b}}, \apj, 803, 34

\bibitem[{{Bruzual} \& {Charlot}(2003)}]{Bruzual:2003aa}
{Bruzual}, G., \& {Charlot}, S. 2003, \mnras, 344, 1000

\bibitem[{{Calzetti} {et~al.}(2000){Calzetti}, {Armus}, {Bohlin}, {Kinney},
  {Koornneef}, \& {Storchi-Bergmann}}]{Calzetti:2000aa}
{Calzetti}, D., {Armus}, L., {Bohlin}, R.~C., {et~al.} 2000, \apj, 533, 682

\bibitem[{{Cen} \& {Kimm}(2015)}]{Cen:2015aa}
{Cen}, R., \& {Kimm}, T. 2015, \apjl, 801, L25

\bibitem[{{Cooke} {et~al.}(2014){Cooke}, {Ryan-Weber}, {Garel}, \&
  {D{\'{\i}}az}}]{Cooke:2014aa}
{Cooke}, J., {Ryan-Weber}, E.~V., {Garel}, T., \& {D{\'{\i}}az}, C.~G. 2014,
  \mnras, 441, 837

\bibitem[{{Faisst} {et~al.}(2016){Faisst}, {Capak}, {Hsieh}, {Laigle},
  {Salvato}, {Tasca}, {Cassata}, {Davidzon}, {Ilbert}, {Le F{\`e}vre},
  {Masters}, {McCracken}, {Steinhardt}, {Silverman}, {de Barros}, {Hasinger},
  \& {Scoville}}]{Faisst:2016aa}
{Faisst}, A.~L., {Capak}, P., {Hsieh}, B.~C., {et~al.} 2016, \apj, 821, 122

\bibitem[{{Fan} {et~al.}(2006{\natexlab{a}}){Fan}, {Carilli}, \&
  {Keating}}]{Fan:2006lr}
{Fan}, X., {Carilli}, C.~L., \& {Keating}, B. 2006{\natexlab{a}}, \araa, 44,
  415

\bibitem[{{Fan} {et~al.}(2006{\natexlab{b}}){Fan}, {Strauss}, {Becker},
  {White}, {Gunn}, {Knapp}, {Richards}, {Schneider}, {Brinkmann}, \&
  {Fukugita}}]{Fan:2006aa}
{Fan}, X., {Strauss}, M.~A., {Becker}, R.~H., {et~al.} 2006{\natexlab{b}}, \aj,
  132, 117

\bibitem[{{Heckman} {et~al.}(2011){Heckman}, {Borthakur}, {Overzier},
  {Kauffmann}, {Basu-Zych}, {Leitherer}, {Sembach}, {Martin}, {Rich},
  {Schiminovich}, \& {Seibert}}]{Heckman:2011aa}
{Heckman}, T.~M., {Borthakur}, S., {Overzier}, R., {et~al.} 2011, \apj, 730, 5

\bibitem[{{Inoue} \& {Iwata}(2008)}]{Inoue:2008aa}
{Inoue}, A.~K., \& {Iwata}, I. 2008, \mnras, 387, 1681

\bibitem[{{Izotov} {et~al.}(2016{\natexlab{a}}){Izotov}, {Orlitov{\'a}},
  {Schaerer}, {Thuan}, {Verhamme}, {Guseva}, \& {Worseck}}]{Izotov:2016aa}
{Izotov}, Y.~I., {Orlitov{\'a}}, I., {Schaerer}, D., {et~al.}
  2016{\natexlab{a}}, \nat, 529, 178

\bibitem[{{Izotov} {et~al.}(2016{\natexlab{b}}){Izotov}, {Schaerer}, {Thuan},
  {Worseck}, {Guseva}, {Orlitov{\'a}}, \& {Verhamme}}]{Izotov:2016ab}
{Izotov}, Y.~I., {Schaerer}, D., {Thuan}, T.~X., {et~al.} 2016{\natexlab{b}},
  \mnras, 461, 3683

\bibitem[{{Kimm} \& {Cen}(2014)}]{Kimm:2014aa}
{Kimm}, T., \& {Cen}, R. 2014, \apj, 788, 121

\bibitem[{{Leitet} {et~al.}(2013){Leitet}, {Bergvall}, {Hayes}, {Linn{\'e}}, \&
  {Zackrisson}}]{Leitet:2013aa}
{Leitet}, E., {Bergvall}, N., {Hayes}, M., {Linn{\'e}}, S., \& {Zackrisson}, E.
  2013, \aap, 553, A106

\bibitem[{{Leitherer} {et~al.}(2016){Leitherer}, {Hernandez}, {Lee}, \&
  {Oey}}]{Leitherer:2016aa}
{Leitherer}, C., {Hernandez}, S., {Lee}, J.~C., \& {Oey}, M.~S. 2016, \apj,
  823, 64

\bibitem[{{Livermore} {et~al.}(2017){Livermore}, {Finkelstein}, \&
  {Lotz}}]{Livermore:2017aa}
{Livermore}, R.~C., {Finkelstein}, S.~L., \& {Lotz}, J.~M. 2017, \apj, 835, 113

\bibitem[{{Madau}(1995)}]{Madau:1995lr}
{Madau}, P. 1995, \apj, 441, 18

\bibitem[{{Naidu} {et~al.}(2016){Naidu}, {Oesch}, {Reddy}, {Holden}, {Steidel},
  {Montes}, {Atek}, {Bouwens}, {Carollo}, {Cibinel}, {Illingworth}, {Labbe},
  {Magee}, {Morselli}, {Nelson}, {van Dokkum}, \& {Wilkins}}]{Naidu:2016aa}
{Naidu}, R.~P., {Oesch}, P.~A., {Reddy}, N., {et~al.} 2016, ArXiv e-prints,
  arXiv:1611.07038

\bibitem[{{Nakajima} \& {Ouchi}(2014)}]{Nakajima:2014aa}
{Nakajima}, K., \& {Ouchi}, M. 2014, \mnras, 442, 900

\bibitem[{{Nestor} {et~al.}(2013){Nestor}, {Shapley}, {Kornei}, {Steidel}, \&
  {Siana}}]{Nestor:2013aa}
{Nestor}, D.~B., {Shapley}, A.~E., {Kornei}, K.~A., {Steidel}, C.~C., \&
  {Siana}, B. 2013, \apj, 765, 47

\bibitem[{{Nestor} {et~al.}(2011){Nestor}, {Shapley}, {Steidel}, \&
  {Siana}}]{Nestor:2011ab}
{Nestor}, D.~B., {Shapley}, A.~E., {Steidel}, C.~C., \& {Siana}, B. 2011, \apj,
  736, 18

\bibitem[{{Ouchi} {et~al.}(2009){Ouchi}, {Mobasher}, {Shimasaku}, {Ferguson},
  {Fall}, {Ono}, {Kashikawa}, {Morokuma}, {Nakajima}, {Okamura}, {Dickinson},
  {Giavalisco}, \& {Ohta}}]{Ouchi:2009aa}
{Ouchi}, M., {Mobasher}, B., {Shimasaku}, K., {et~al.} 2009, \apj, 706, 1136

\bibitem[{{Pawlik} {et~al.}(2009){Pawlik}, {Schaye}, \& {van
  Scherpenzeel}}]{Pawlik:2009aa}
{Pawlik}, A.~H., {Schaye}, J., \& {van Scherpenzeel}, E. 2009, \mnras, 394,
  1812

\bibitem[{{Pettini} \& {Pagel}(2004)}]{Pettini:2004qe}
{Pettini}, M., \& {Pagel}, B.~E.~J. 2004, \mnras, 348, L59

\bibitem[{{Planck Collaboration} {et~al.}(2016){Planck Collaboration}, {Adam},
  {Aghanim}, {Ashdown}, {Aumont}, {Baccigalupi}, {Ballardini}, {Banday},
  {Barreiro}, {Bartolo}, {Basak}, {Battye}, {Benabed}, {Bernard}, {Bersanelli},
  {Bielewicz}, {Bock}, {Bonaldi}, {Bonavera}, {Bond}, {Borrill}, {Bouchet},
  {Boulanger}, {Bucher}, {Burigana}, {Calabrese}, {Cardoso}, {Carron},
  {Chiang}, {Colombo}, {Combet}, {Comis}, {Couchot}, {Coulais}, {Crill},
  {Curto}, {Cuttaia}, {Davis}, {de Bernardis}, {de Rosa}, {de Zotti},
  {Delabrouille}, {Di Valentino}, {Dickinson}, {Diego}, {Dor{\'e}}, {Douspis},
  {Ducout}, {Dupac}, {Elsner}, {En{\ss}lin}, {Eriksen}, {Falgarone}, {Fantaye},
  {Finelli}, {Forastieri}, {Frailis}, {Fraisse}, {Franceschi}, {Frolov},
  {Galeotta}, {Galli}, {Ganga}, {G{\'e}nova-Santos}, {Gerbino}, {Ghosh},
  {Gonz{\'a}lez-Nuevo}, {G{\'o}rski}, {Gruppuso}, {Gudmundsson}, {Hansen},
  {Helou}, {Henrot-Versill{\'e}}, {Herranz}, {Hivon}, {Huang}, {Ili{\'c}},
  {Jaffe}, {Jones}, {Keih{\"a}nen}, {Keskitalo}, {Kisner}, {Knox},
  {Krachmalnicoff}, {Kunz}, {Kurki-Suonio}, {Lagache}, {L{\"a}hteenm{\"a}ki},
  {Lamarre}, {Langer}, {Lasenby}, {Lattanzi}, {Lawrence}, {Le Jeune},
  {Levrier}, {Lewis}, {Liguori}, {Lilje}, {L{\'o}pez-Caniego}, {Ma},
  {Mac{\'{\i}}as-P{\'e}rez}, {Maggio}, {Mangilli}, {Maris}, {Martin},
  {Mart{\'{\i}}nez-Gonz{\'a}lez}, {Matarrese}, {Mauri}, {McEwen}, {Meinhold},
  {Melchiorri}, {Mennella}, {Migliaccio}, {Miville-Desch{\^e}nes}, {Molinari},
  {Moneti}, {Montier}, {Morgante}, {Moss}, {Naselsky}, {Natoli}, {Oxborrow},
  {Pagano}, {Paoletti}, {Partridge}, {Patanchon}, {Patrizii}, {Perdereau},
  {Perotto}, {Pettorino}, {Piacentini}, {Plaszczynski}, {Polastri}, {Polenta},
  {Puget}, {Rachen}, {Racine}, {Reinecke}, {Remazeilles}, {Renzi}, {Rocha},
  {Rossetti}, {Roudier}, {Rubi{\~n}o-Mart{\'{\i}}n}, {Ruiz-Granados},
  {Salvati}, {Sandri}, {Savelainen}, {Scott}, {Sirri}, {Sunyaev}, {Suur-Uski},
  {Tauber}, {Tenti}, {Toffolatti}, {Tomasi}, {Tristram}, {Trombetti},
  {Valiviita}, {Van Tent}, {Vielva}, {Villa}, {Vittorio}, {Wandelt}, {Wehus},
  {White}, {Zacchei}, \& {Zonca}}]{Planck-Collaboration:2016aa}
{Planck Collaboration}, {Adam}, R., {Aghanim}, N., {et~al.} 2016, \aap, 596,
  A108

\bibitem[{{Prevot} {et~al.}(1984){Prevot}, {Lequeux}, {Prevot}, {Maurice}, \&
  {Rocca-Volmerange}}]{prevot:1984aa}
{Prevot}, M.~L., {Lequeux}, J., {Prevot}, L., {Maurice}, E., \&
  {Rocca-Volmerange}, B. 1984, \aap, 132, 389

\bibitem[{{Reddy} {et~al.}(2010){Reddy}, {Erb}, {Pettini}, {Steidel}, \&
  {Shapley}}]{Reddy:2010aa}
{Reddy}, N.~A., {Erb}, D.~K., {Pettini}, M., {Steidel}, C.~C., \& {Shapley},
  A.~E. 2010, \apj, 712, 1070

\bibitem[{{Reddy} {et~al.}(2016){Reddy}, {Steidel}, {Pettini},
  {Bogosavljevi{\'c}}, \& {Shapley}}]{Reddy:2016aa}
{Reddy}, N.~A., {Steidel}, C.~C., {Pettini}, M., {Bogosavljevi{\'c}}, M., \&
  {Shapley}, A.~E. 2016, \apj, 828, 108

\bibitem[{{Richard} {et~al.}(2011){Richard}, {Jones}, {Ellis}, {Stark},
  {Livermore}, \& {Swinbank}}]{Richard:2011aa}
{Richard}, J., {Jones}, T., {Ellis}, R., {et~al.} 2011, \mnras, 413, 643

\bibitem[{{Robertson} {et~al.}(2013){Robertson}, {Furlanetto}, {Schneider},
  {Charlot}, {Ellis}, {Stark}, {McLure}, {Dunlop}, {Koekemoer}, {Schenker},
  {Ouchi}, {Ono}, {Curtis-Lake}, {Rogers}, {Bowler}, \&
  {Cirasuolo}}]{Robertson:2013dq}
{Robertson}, B.~E., {Furlanetto}, S.~R., {Schneider}, E., {et~al.} 2013, \apj,
  768, 71

\bibitem[{{Rutkowski} {et~al.}(2016){Rutkowski}, {Scarlata}, {Haardt}, {Siana},
  {Henry}, {Rafelski}, {Hayes}, {Salvato}, {Pahl}, {Mehta}, {Beck}, {Malkan},
  \& {Teplitz}}]{Rutkowski:2016aa}
{Rutkowski}, M.~J., {Scarlata}, C., {Haardt}, F., {et~al.} 2016, \apj, 819, 81

\bibitem[{{Salpeter}(1955)}]{Salpeter:1955aa}
{Salpeter}, E.~E. 1955, \apj, 121, 161

\bibitem[{{Schenker} {et~al.}(2014){Schenker}, {Ellis}, {Konidaris}, \&
  {Stark}}]{Schenker:2014aa}
{Schenker}, M.~A., {Ellis}, R.~S., {Konidaris}, N.~P., \& {Stark}, D.~P. 2014,
  \apj, 795, 20

\bibitem[{{Shapley} {et~al.}(2006){Shapley}, {Steidel}, {Pettini},
  {Adelberger}, \& {Erb}}]{Shapley:2006kx}
{Shapley}, A.~E., {Steidel}, C.~C., {Pettini}, M., {Adelberger}, K.~L., \&
  {Erb}, D.~K. 2006, \apj, 651, 688

\bibitem[{{Shapley} {et~al.}(2016){Shapley}, {Steidel}, {Strom},
  {Bogosavljevi{\'c}}, {Reddy}, {Siana}, {Mostardi}, \&
  {Rudie}}]{Shapley:2016aa}
{Shapley}, A.~E., {Steidel}, C.~C., {Strom}, A.~L., {et~al.} 2016, \apjl, 826,
  L24

\bibitem[{{Siana} {et~al.}(2007){Siana}, {Teplitz}, {Colbert}, {Ferguson},
  {Dickinson}, {Brown}, {Conselice}, {de Mello}, {Gardner}, {Giavalisco}, \&
  {Menanteau}}]{Siana:2007lr}
{Siana}, B., {Teplitz}, H.~I., {Colbert}, J., {et~al.} 2007, \apj, 668, 62

\bibitem[{{Siana} {et~al.}(2015){Siana}, {Shapley}, {Kulas}, {Nestor},
  {Steidel}, {Teplitz}, {Alavi}, {Brown}, {Conselice}, {Ferguson}, {Dickinson},
  {Giavalisco}, {Colbert}, {Bridge}, {Gardner}, \& {de Mello}}]{Siana:2015aa}
{Siana}, B., {Shapley}, A.~E., {Kulas}, K.~R., {et~al.} 2015, \apj, 804, 17

\bibitem[{{Stark} {et~al.}(2011){Stark}, {Ellis}, \& {Ouchi}}]{Stark:2011fk}
{Stark}, D.~P., {Ellis}, R.~S., \& {Ouchi}, M. 2011, \apjl, 728, L2

\bibitem[{{Steidel} {et~al.}(2001){Steidel}, {Pettini}, \&
  {Adelberger}}]{Steidel:2001aa}
{Steidel}, C.~C., {Pettini}, M., \& {Adelberger}, K.~L. 2001, \apj, 546, 665

\bibitem[{{Vanzella} {et~al.}(2010){Vanzella}, {Giavalisco}, {Inoue}, {Nonino},
  {Fontanot}, {Cristiani}, {Grazian}, {Dickinson}, {Stern}, {Tozzi},
  {Giallongo}, {Ferguson}, {Spinrad}, {Boutsia}, {Fontana}, {Rosati}, \&
  {Pentericci}}]{Vanzella:2010aa}
{Vanzella}, E., {Giavalisco}, M., {Inoue}, A.~K., {et~al.} 2010, \apj, 725,
  1011

\bibitem[{{Vanzella} {et~al.}(2012){Vanzella}, {Guo}, {Giavalisco}, {Grazian},
  {Castellano}, {Cristiani}, {Dickinson}, {Fontana}, {Nonino}, {Giallongo},
  {Pentericci}, {Galametz}, {Faber}, {Ferguson}, {Grogin}, {Koekemoer},
  {Newman}, \& {Siana}}]{Vanzella:2012aa}
{Vanzella}, E., {Guo}, Y., {Giavalisco}, M., {et~al.} 2012, \apj, 751, 70

\bibitem[{{Vanzella} {et~al.}(2016){Vanzella}, {de Barros}, {Vasei}, {Alavi},
  {Giavalisco}, {Siana}, {Grazian}, {Hasinger}, {Suh}, {Cappelluti}, {Vito},
  {Amorin}, {Balestra}, {Brusa}, {Calura}, {Castellano}, {Comastri}, {Fontana},
  {Gilli}, {Mignoli}, {Pentericci}, {Vignali}, \& {Zamorani}}]{Vanzella:2016ab}
{Vanzella}, E., {de Barros}, S., {Vasei}, K., {et~al.} 2016, \apj, 825, 41

\bibitem[{{Vasei} {et~al.}(2016){Vasei}, {Siana}, {Shapley}, {Quider}, {Alavi},
  {Rafelski}, {Steidel}, {Pettini}, \& {Lewis}}]{Vasei:2016aa}
{Vasei}, K., {Siana}, B., {Shapley}, A.~E., {et~al.} 2016, \apj, 831, 38

\bibitem[{{Verhamme} {et~al.}(2017){Verhamme}, {Orlitov{\'a}}, {Schaerer},
  {Izotov}, {Worseck}, {Thuan}, \& {Guseva}}]{Verhamme:2017aa}
{Verhamme}, A., {Orlitov{\'a}}, I., {Schaerer}, D., {et~al.} 2017, \aap, 597,
  A13

\end{thebibliography}
\end{document}